\documentclass[conference]{IEEEtran}

\usepackage{cite}
\usepackage{amsmath,amssymb,amsfonts}
\usepackage{algorithmic}
\usepackage{graphicx}
\usepackage{textcomp}
\usepackage{xcolor}
\usepackage{nomencl}
\usepackage[numbers]{natbib}
\usepackage{etoolbox}

\AtBeginEnvironment{subequations}{%
    \setlength{\abovedisplayskip}{2pt}%
    \setlength{\belowdisplayskip}{2pt}%
    \setlength{\abovedisplayshortskip}{0pt}%
    \setlength{\belowdisplayshortskip}{0pt}%
}

\def\BibTeX{{\rm B\kern-.05em{\sc i\kern-.025em b}\kern-.08em
    T\kern-.1667em\lower.7ex\hbox{E}\kern-.125emX}}
\begin{document}
\addtolength{\topmargin}{0.03in}
\addtolength{\textheight}{-0.03in}

\makenomenclature

\renewcommand{\nomgroup}[1]{%
  \ifthenelse{\equal{#1}{A}}{\item[\textbf{Sets and Indices}]}{}%
  \ifthenelse{\equal{#1}{B}}{\item[\textbf{Parameters}]}{}%
  \ifthenelse{\equal{#1}{C}}{\item[\textbf{Variables}]}{}%
}
\setlength{\nomitemsep}{1pt} 


\title{Massachusetts' 2026 Clean Peak Standard Recalibration: Adaptation and Storage Tradeoffs}

\author{
\IEEEauthorblockN{Yuhan Dai}
\IEEEauthorblockA{\textit{Thayer School of Engineering} \\
\textit{Dartmouth College}\\
Hanover, NH, US \\
yuhan.dai.th@dartmouth.edu}
\and
\IEEEauthorblockN{Boyu Yao}
\IEEEauthorblockA{\textit{Department of Civil and Systems Engineering} \\
\textit{Johns Hopkins University}\\
Baltimore, MD, US \\
byao3@jhu.edu}
}

\maketitle

\begin{abstract}
Massachusetts recalibrated its Clean Peak Standard (CPS) in 2026 by lowering the minimum standards and expanding the Near-Term Resource Multiplier for qualifying storage. This paper evaluates the change using a three-zone, hourly capacity expansion model for 2026--2030. Both scenarios achieve full CPS compliance, but the \emph{post-May} scenario reduces new battery additions by 53.1\% and modeled system cost by 3.7\%. CPS-eligible discharge declines by 30.4\%, while Clean Peak Energy Certificate production declines by only 9.5\%. Fossil generation during CPS hours increases by 11.7\%, whereas modeled emissions remain nearly unchanged because the Regional Greenhouse Gas Initiative cap binds. The recalibration therefore lowers the modeled storage capacity requirement while weakening physical clean-peak performance.
\end{abstract}


{\small
\printnomenclature
}


\nomenclature[A01]{$\mathcal{I}$}
{Set of Massachusetts zones; indexed by $i,j$.}%

\nomenclature[A02]{$\mathcal{L}/\mathcal{LS}_{i}/\mathcal{LR}_{i}$}
{Set of transmission corridors and subsets sending from and receiving into zone $i$; indexed by $l$.}%

\nomenclature[A03]{$\mathcal{K}/\mathcal{N}$}
{Set of technologies and resource types; indexed by $k$ and $n$, respectively.}%

\nomenclature[A04]{$\mathcal{G}/\mathcal{G}_{i}/\mathcal{G}^{n}/
\mathcal{G}^{k}/\mathcal{G}^{\mathrm{RGGI}}/
\bar{\mathcal{G}}/\tilde{\mathcal{G}}$}
{Set of units and subsets by zone, type, technology, RGGI coverage, existing status, and candidate status; indexed by $g$.}%

\nomenclature[A05]{$\mathcal{Y}$}
{Set of planning years; indexed by $y$.}%

\nomenclature[A06]{$\mathcal{T}/\mathcal{H}$}
{Set of representative days and hours; indexed by $t$ and $h$.}%



\nomenclature[B01]{$L_{ithy}$}
{Net load in zone $i$, day $t$, hour $h$, and year $y$, after subtracting assumed imports (MW).}%

\nomenclature[B02]{$\overline{L}_{y}$}
{Coincident Massachusetts net peak load in year $y$ (MW).}%

\nomenclature[B03]{$\overline{P}^{\mathrm{G}}_{g}/
\overline{P}^{\mathrm{L},+}_{l}/
\overline{P}^{\mathrm{L},-}_{l}$}
{Capacity of unit $g$ and forward and reverse transfer limits of corridor $l$ (MW).}%

\nomenclature[B04]{$F^{\mathrm{GEN}}_{igthy}$}
{Renewable availability factor for unit $g$ in zone $i$, day $t$, hour $h$, and year $y$ (unitless).}%

\nomenclature[B05]{$\overline{E}^{\mathrm{ST}}_{g}$}
{Energy capacity of storage unit $g$ (MWh).}%

\nomenclature[B06]{$\epsilon^{\mathrm{CH}}_{g}/
\epsilon^{\mathrm{DC}}_{g}$}
{Charging and discharging efficiencies of storage unit $g$ (unitless).}%

\nomenclature[B07]{$F^{\mathrm{ELCC}}_{ky}$}
{ELCC factor of technology $k$ in year $y$ (unitless).}%

\nomenclature[B08]{$R^{\mathrm{RA}}_{y}$}
{Resource-adequacy planning reserve in year $y$ (unitless).}%

\nomenclature[B09]{$R^{\mathrm{CPS}}_{y}$}
{CPS minimum compliance percentage in year $y$ (unitless).}%

\nomenclature[B10]{$M^{\mathrm{CPS}}_{thy}$}
{Hourly CPS multiplier (unitless).}%

\nomenclature[B11]{$M^{\mathrm{NTRM}}_{gy}$}
{Near-Term Resource Multiplier for storage unit $g$ in year $y$ (unitless).}%

\nomenclature[B12]{$\epsilon^{\mathrm{CO_2}}_{g}$}
{CO$_2$ emissions rate of unit $g$ (tons/MWh).}%

\nomenclature[B13]{$\overline{E}^{\mathrm{CO_2}}_y$}
{Annual upper limit on CO$_2$ emissions in year $y$ (tons).}%

\nomenclature[B14]{$P^{\mathrm{VOLL}}$}
{Value of Lost Load (\$/MWh).}%

\nomenclature[B15]{$P^{\mathrm{RA}}$}
{Penalty for resource-adequacy shortfall (\$/MW-year).}%

\nomenclature[B16]{$P^{\mathrm{CPS}}_{y}$}
{CPS Alternative Compliance Payment in year $y$ (\$/CPEC).}%

\nomenclature[B17]{$P^{\mathrm{CO_2}}_{y}$}
{RGGI allowance price in year $y$ (\$/short ton of CO$_2$).}%

\nomenclature[B18]{$C^{\mathrm{I}}_{gy}$}
{Capital investment cost of unit $g$ in year $y$ (\$/MW).}%

\nomenclature[B19]{$C^{\mathrm{F}}_{g}/C^{\mathrm{V}}_{g}$}
{Fixed and variable operating costs of unit $g$ (\$/MW-year and \$/MWh).}%

\nomenclature[B20]{$N_{ty}$}
{Number of days represented by representative day $t$ in year $y$ (days).}%


\nomenclature[C01]{$p_{gthy}$}
{Power generation of non-storage unit $g$ in day $t$, hour $h$, and year $y$ (MW).}%

\nomenclature[C02]{$q_{lthy}$}
{Power flow of corridor $l$ in day $t$, hour $h$, and year $y$ (MW).}%

\nomenclature[C03]{$p^{\mathrm{LS}}_{ithy}$}
{Load shedding in zone $i$, day $t$, hour $h$, and year $y$ (MW).}%

\nomenclature[C04]{$c_{gthy}/dc_{gthy}$}
{Charging and discharging power of storage unit $g$ in day $t$, hour $h$, and year $y$ (MW).}%

\nomenclature[C05]{$e^{\mathrm{SOC}}_{gthy}$}
{State of charge of storage unit $g$ in day $t$, hour $h$, and year $y$ (MWh).}%

\nomenclature[C06]{$p^{\mathrm{RA}}_{y}$}
{Resource-adequacy shortfall in year $y$ (MW).}%

\nomenclature[C07]{$e^{\mathrm{CPS}}_{y}$}
{CPS compliance shortfall in year $y$ (CPEC).}%

\nomenclature[C08]{$b_{gy}$}
{Investment decision for candidate unit $g$ in year $y$ (unitless).}%

\nomenclature[C09]{$o_{gy}$}
{Operational status of unit $g$ in year $y$ (unitless).}%

\section{Introduction}
\label{sec:introduction}

Decarbonizing electricity requires more than producing sufficient clean energy over the course of a year; clean resources must also be available when demand and the marginal value of supply are highest \cite{davis2018net}. As solar and wind penetration increases, periods of abundant renewable output may not coincide with evening peaks, winter stress events, or hours in which fossil generation sets system emissions. Energy storage can shift electricity across time, reduce renewable curtailment, and provide capacity and flexibility, but its system value depends strongly on when and where it charges and discharges \cite{dunn2011electrical,arbabzadeh2019role}. Profitable storage arbitrage may also diverge from emissions-minimizing or peak-reducing operation when prices do not fully reflect environmental and reliability externalities \cite{hittinger2015bulk}. This potential divergence motivates policies that differentiate the value of clean electricity across time rather than relying only on annual certificate accounting \cite{xu2024system}.

Massachusetts provides a timely policy experiment in this transition from annual clean-energy quantity toward time-sensitive clean-energy value. Its Clean Peak Standard (CPS) rewards eligible generation, storage discharge, and demand reduction during periods of high net electricity demand \cite{mass_doer_cps}. Eligible resources earn Clean Peak Energy Certificates (CPECs), while retail electricity suppliers must acquire enough certificates to satisfy an annual Minimum Standard or cover any shortfall through Alternative Compliance Payments (ACP). Seasonal, hourly, and monthly-peak multipliers increase certificate awards during higher-value periods. CPEC revenues can supplement energy, capacity, and ancillary-service revenues for storage developers, but certificate scarcity can also increase compliance costs for suppliers and consumers.

In May 2026, the Massachusetts Department of Energy Resources adopted emergency amendments that reduced the CPS Minimum Standards relative to the pre-May policy trajectory, as summarized in Table~\ref{tab:cps-scenarios}. The amendments also extended NTRM eligibility for qualifying projects affected by permitting, interconnection, procurement, or construction delays and established December 31, 2036, as the multiplier expiration date \cite{mass_225_cmr_21}. These changes occurred amid broader recognition that clean-energy deployment schedules may be constrained not only by project economics and permitting, but also by material availability, manufacturing capacity, and feasible installation rates \cite{yao2025understanding}. The Department framed the amendments as a means of reducing near-term ratepayer exposure while preserving a pathway for additional storage deployment. The recalibration can therefore be interpreted as an attempt to align near-term certificate obligations with the feasible pace of project delivery without abandoning the CPS's longer-term objectives.

\begin{table}[b]
\vspace{-1em}
\caption{CPS Minimum Standard Trajectories Used in the Model}
\begin{center}
\begin{tabular}{|c|c|c|}
\hline
\textbf{Year}
& \textbf{Pre-May 2026 Policy}
& \textbf{Post-May 2026 Amendments} \\
\hline
2026 & 7\%  & 4\%  \\
\hline
2027 & 9\%  & 8\%  \\
\hline
2028 & 13\% & 12\% \\
\hline
2029 & 19\% & 18\% \\
\hline
2030 & 25\% & 24\% \\
\hline
\end{tabular}
\label{tab:cps-scenarios}
\end{center}

\end{table}

The recalibration creates a policy-design tradeoff. Lower minimum standards reduce CPEC demand and near-term compliance costs, while extended NTRM eligibility increases certificate production from qualifying projects. However, a multiplier awards more certificates per megawatt-hour of discharge without directly increasing battery capacity, physical discharge, or fossil generation displaced during clean peak periods. Compliance may therefore be maintained even as physical storage deployment declines. The central question is whether the revised policy can serve as a strategic bridge through near-term deployment constraints while preserving the system benefits that motivated the CPS.

Existing studies highlight that clean-energy certificate compliance and physical power-system outcomes need not coincide. Shrader et al.\ evaluate the original Massachusetts CPS and show that clean peak incentives can alter battery dispatch without necessarily producing proportional incremental emissions reductions \cite{shrader2021not}. More broadly, Kim et al.\ demonstrate using an ISO New England (ISO-NE) test system that renewable-support policies involve tradeoffs among compliance costs, investment incentives, and emissions outcomes \cite{kim2021strategic}. Griffiths finds that CPS can impose additional system costs without necessarily delivering proportional emissions benefits, highlighting the sensitivity of their effectiveness to policy design and interactions with other clean-energy requirements \cite{griffiths2020expensive}. These findings are directly relevant to the Massachusetts recalibration, which simultaneously lowers system-wide CPEC obligations and expands multiplier eligibility. However, prior work does not evaluate how this combined change affects endogenous battery expansion, physical clean peak discharge, peak-period fossil generation, and total system cost. This study addresses that gap through an hourly capacity expansion analysis of the pre- and post-May 2026 policy pathways.

Accordingly, this paper investigates how Massachusetts' 2026 CPS recalibration affects storage deployment and system performance relative to the pre-May policy trajectory. We use a three-zone, hourly capacity expansion and economic dispatch model for 2026--2030 to compare a counterfactual continuation of the pre-May policy with the May 2026 policy package. The analysis evaluates battery investment, physical clean-peak discharge, CPEC production, peak-period fossil generation, carbon emissions, resource adequacy, and total system cost.

The model does not explicitly represent supply-chain or interconnection delays; rather, it evaluates the system consequences of the policy response motivated by these constraints. The study provides an early quantitative assessment of the recalibrated CPS and distinguishes certificate compliance from physical clean-peak performance, thereby identifying the tradeoff between near-term policy feasibility and continued storage-deployment incentives.

\section{Study Scope and Model Description}
\label{sec:model}

\begin{figure}[b]
\vspace{-2em}
    \centering
    \includegraphics[width=0.5\textwidth]{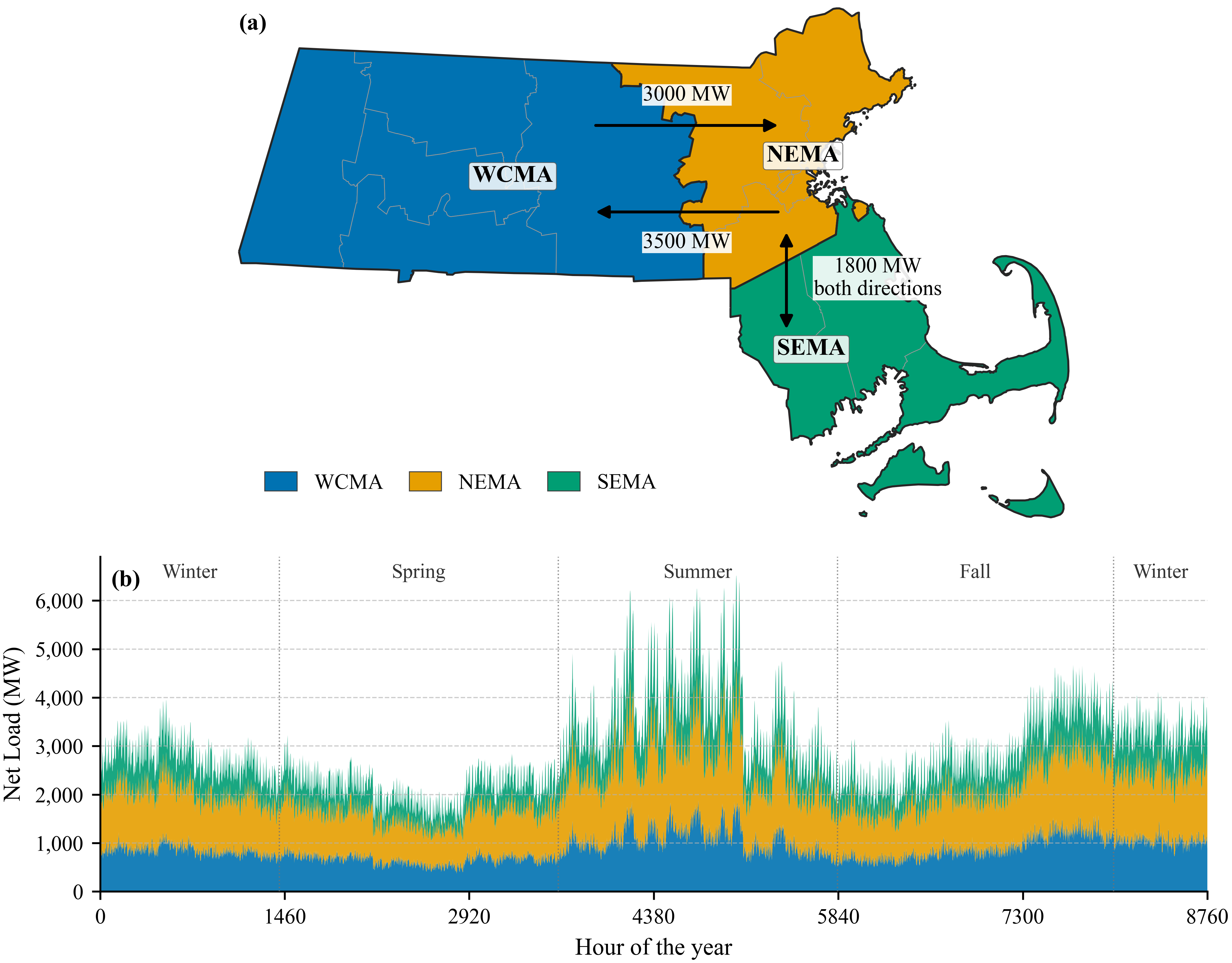}
    \vspace{-1em}
    \caption{Modeled Massachusetts power system: (a) three-zone network and transfer limits; and (b) zonal net load profile for all 8760 hours of 2026.}
    \label{fig:system_rep}
\end{figure}
 
\subsection{Study Scope and Policy Scenarios}
\label{subsec:scope}

We formulate a multi-stage capacity expansion and economic dispatch model of the Massachusetts power system over 2026--2030. As shown in Fig.~\ref{fig:system_rep}, the system is represented by three zones: Northeastern Massachusetts (NEMA), Southeastern Massachusetts (SEMA), and Western and Central Massachusetts (WCMA). The model jointly determines generation and battery storage investments and hourly system operation over the full 8760-hour chronology of each planning year. Electricity demand, interzonal transfer limits, resource-adequacy requirements, the CPS, and the Regional Greenhouse Gas Initiative (RGGI) emissions cap are enforced. Existing  resources available in 2026 are treated as fixed capacity, while candidate investments are selected endogenously to minimize total investment, operating, and policy-compliance costs over the planning horizon.

The analysis compares two CPS policy scenarios. The \emph{pre-May} scenario uses the CPS minimum standards in effect before the May 2026 amendments and assigns a Near-Term Resource Multiplier\footnote{The NTRM increases the number of CPECs
produced by each eligible MWh discharged from qualifying new storage.} (NTRM) of one to candidate batteries. The \emph{post-May} scenario uses the revised minimum standards and assigns a $2 \times$ NTRM to qualifying candidate batteries. The annual minimum standards are listed in
Table~\ref{tab:cps-scenarios}.

In the \emph{post-May} scenario, all modeled candidate batteries are assumed to qualify for the $2 \times$ NTRM. This assumption represents the amended policy as a common treatment for eligible new projects. The results therefore show the combined effect of the revised minimum standards and NTRM treatment.

\subsection{Multi-Stage Capacity Expansion Model}
\label{subsec:capacity-expansion}

\subsubsection{Objective Function}
\label{subsubsec:objective}

The objective in \eqref{eq:objective} minimizes total investment, operating, and policy-related costs over the planning horizon. Investment cost \(C^{\mathrm{in}}_y\), defined in \eqref{eq:investment_cost}, represents the annualized cost of candidate capacity available in year \(y\), evaluated using the adjusted capital-cost coefficient \(AC^{\mathrm{I}}_{gy}\)\footnote{\(AC^{\mathrm{I}}_{gy}\) denotes the annualized capital cost of candidate unit \(g\), adjusted for its effective operating years within the planning horizon.}. Operating cost \(C^{\mathrm{op}}_y\) in \eqref{eq:operation_cost} includes fixed O\&M costs, variable generation costs, and battery charging and discharging costs. Policy and penalty cost \(C^{\mathrm{pe}}_y\) in \eqref{eq:penalty_cost} includes unserved-energy costs, resource adequacy and CPS shortfall penalties, and RGGI allowance payments. All monetary terms are expressed in consistent annual dollars and summed over the modeled years.

\begin{subequations}
\label{eq:objective_group}

\begin{equation}
\min
\sum_{y\in\mathcal{Y}}
\left(
C^{\mathrm{in}}_y
+
C^{\mathrm{op}}_y
+
C^{\mathrm{pe}}_y
\right)
\label{eq:objective}
\end{equation}

\begin{equation}
C^{\mathrm{in}}_y
=
\sum_{g\in\tilde{\mathcal{G}}}
AC^{\mathrm{I}}_{gy}
\overline{P}^{\mathrm{G}}_g
o_{gy}
\label{eq:investment_cost}
\end{equation}

\begin{equation}
\begin{aligned}
C^{\mathrm{op}}_y
={}&
\sum_{g\in\mathcal{G}}
C^{\mathrm{F}}_g
\overline{P}^{\mathrm{G}}_g
o_{gy}
\\
&+
\sum_{g\in\mathcal{G}^{\mathrm{th}}\cup\mathcal{G}^{\mathrm{rn}}}
\sum_{t\in\mathcal{T}}
\sum_{h\in\mathcal{H}}
C^{\mathrm{V}}_g
N_{ty}
p_{gthy}
\\
&+
\sum_{g\in\mathcal{G}^{\mathrm{st}}}
\sum_{t\in\mathcal{T}}
\sum_{h\in\mathcal{H}}
C^{\mathrm{V}}_g
N_{ty}
\left(
c_{gthy}+dc_{gthy}
\right)
\end{aligned}
\label{eq:operation_cost}
\end{equation}

\begin{equation}
\begin{aligned}
C^{\mathrm{pe}}_y
={}&
P^{\mathrm{VOLL}}
\sum_{i\in\mathcal{I}}
\sum_{t\in\mathcal{T}}
\sum_{h\in\mathcal{H}}
N_{ty}
p^{\mathrm{LS}}_{ithy}
+
P^{\mathrm{RA}}p^{\mathrm{RA}}_y
+
P^{\mathrm{CPS}}_y e^{\mathrm{CPS}}_y
\\
&+
P^{\mathrm{CO_2}}_y
\sum_{g\in\mathcal{G}^{\mathrm{RGGI}}}
\sum_{t\in\mathcal{T}}
\sum_{h\in\mathcal{H}}
N_{ty}
\epsilon^{\mathrm{CO_2}}_g
p_{gthy}
\end{aligned}
\label{eq:penalty_cost}
\end{equation}

\end{subequations}

\subsubsection{System Operation and Expansion Constraints}
\label{subsubsec:system-constraints}

\begin{subequations}
\label{eq:system_constraints}

\begin{equation}
\begin{aligned}
&
\sum_{g\in\mathcal{G}_{i}\cap
(\mathcal{G}^{\mathrm{th}}\cup\mathcal{G}^{\mathrm{rn}})}
p_{gthy}
+
\sum_{g\in\mathcal{G}_{i}\cap\mathcal{G}^{\mathrm{st}}}
\left(dc_{gthy}-c_{gthy}\right)
\\
&\quad
-\sum_{l\in\mathcal{LS}_{i}}q_{lthy}
+\sum_{l\in\mathcal{LR}_{i}}q_{lthy}
=
L_{ithy}-p^{\mathrm{LS}}_{ithy},
\qquad
\forall i,t,h,y
\end{aligned}
\label{eq:energy_balance}
\end{equation}

\begin{equation}
0\leq p_{gthy}
\leq
\overline{P}^{\mathrm{G}}_{g}o_{gy},
\qquad
\forall g\in\mathcal{G}^{\mathrm{th}},\ t,h,y
\label{eq:thermal_output}
\end{equation}

\begin{equation}
0\leq p_{gthy}
\leq
F^{\mathrm{GEN}}_{igthy}
\overline{P}^{\mathrm{G}}_{g}o_{gy},
\qquad
\forall g\in\mathcal{G}_{i}\cap\mathcal{G}^{\mathrm{rn}},\ i,t,h,y
\label{eq:renewable_output}
\end{equation}

\begin{equation}
o_{g1}=b_{g1},
\qquad
o_{gy}=o_{g(y-1)}+b_{gy},
\quad
\forall g\in\tilde{\mathcal{G}},\ y\geq 2
\label{eq:operational_status}
\end{equation}

\begin{equation}
0\leq b_{gy}\leq 1,
\quad
\sum_{y\in\mathcal{Y}} b_{gy}\leq 1,
\quad
0\leq o_{gy}\leq 1,
\quad
\forall g\in\tilde{\mathcal{G}},\ y
\label{eq:candidate_status}
\end{equation}

\begin{equation}
o_{gy}=1,
\qquad
\forall g\in\bar{\mathcal{G}},\ y
\label{eq:existing_status}
\end{equation}

\begin{equation}
\overline{P}^{\mathrm{L},-}_{l}
\leq q_{lthy}
\leq
\overline{P}^{\mathrm{L},+}_{l},
\qquad
\forall l,t,h,y
\label{eq:transmission}
\end{equation}

\begin{equation}
0\leq c_{gthy}
\leq
\overline{P}^{\mathrm{G}}_{g}o_{gy},
\qquad
\forall g\in\mathcal{G}^{\mathrm{st}},\ t,h,y
\label{eq:storage_charging}
\end{equation}

\begin{equation}
0\leq dc_{gthy}
\leq
\overline{P}^{\mathrm{G}}_{g}o_{gy},
\qquad
\forall g\in\mathcal{G}^{\mathrm{st}},\ t,h,y
\label{eq:storage_discharging}
\end{equation}

\begin{equation}
0\leq e^{\mathrm{SOC}}_{gthy}
\leq
\overline{E}^{\mathrm{ST}}_{g}o_{gy},
\qquad
\forall g\in\mathcal{G}^{\mathrm{st}},\ t,h,y
\label{eq:soc_limit}
\end{equation}

\begin{equation}
e^{\mathrm{SOC}}_{gthy}
=
e^{\mathrm{SOC}}_{g,\pi(t,h),y}
+
\epsilon^{\mathrm{CH}}_{g}c_{gthy}
-
\frac{dc_{gthy}}{\epsilon^{\mathrm{DC}}_{g}},
\quad
\forall g\in\mathcal{G}^{\mathrm{st}},\ t,h,y
\label{eq:storage_dynamics}
\end{equation}

\begin{equation}
0\leq p^{\mathrm{LS}}_{ithy}\leq L_{ithy},
\qquad
\forall i,t,h,y
\label{eq:load_shedding}
\end{equation}

\end{subequations}

Eq.~\eqref{eq:energy_balance} enforces the hourly zonal power balance,
including generation, storage operation, interzonal transfers, and load
shedding. Eqs.~\eqref{eq:thermal_output} and \eqref{eq:renewable_output}
limit thermal and renewable output by available capacity, with renewable
generation further restricted by hourly availability. Eqs.~\eqref{eq:operational_status}--%
\eqref{eq:existing_status} track cumulative candidate additions and preserve
existing capacity, while Eq.~\eqref{eq:transmission} imposes directional
interzonal transfer limits.

Eqs.~\eqref{eq:storage_charging}--\eqref{eq:storage_dynamics} define storage
power limits, energy capacity, and chronological state-of-charge evolution.
Here, \(\pi(t,h)\) denotes the immediately preceding operating period,
including the transition between consecutive days. State of charge is linked
across all 8760 hours of each model year, with initial and terminal levels
fixed at 50\% of available energy capacity. Finally,
Eq.~\eqref{eq:load_shedding} bounds unserved energy by zonal load.

\subsubsection{Massachusetts Policy Compliance Constraints}
\label{subsubsec:policy-constraints}

\begin{subequations}
\label{eq:policy_constraints}

\begin{equation}
\sum_{k\in\mathcal{K}}
\sum_{g\in\mathcal{G}^{k}}
\overline{P}^{\mathrm{G}}_{g}
F^{\mathrm{ELCC}}_{ky}
o_{gy}
+
p^{\mathrm{RA}}_{y}
\geq
\left(1+R^{\mathrm{RA}}_{y}\right)\overline{L}_{y},
\qquad
\forall y
\label{eq:resource_adequacy}
\end{equation}

\begin{equation}
\sum_{g\in\mathcal{G}^{\mathrm{th}}}
\sum_{t\in\mathcal{T}}
\sum_{h\in\mathcal{H}}
N_{ty}
\epsilon^{\mathrm{CO_2}}_{g}
p_{gthy}
\leq
\overline{E}^{\mathrm{CO_2}}_{y},
\qquad
\forall y
\label{eq:co2_cap}
\end{equation}

\begin{equation}
\begin{aligned}
&
\sum_{g\in\mathcal{G}^{\mathrm{BSS}}}
\sum_{t\in\mathcal{T}}
\sum_{h\in\mathcal{H}}
N_{ty}
M^{\mathrm{CPS}}_{thy}
M^{\mathrm{NTRM}}_{gy}
dc_{gthy}
+
e^{\mathrm{CPS}}_{y}
\\
&\qquad\geq
R^{\mathrm{CPS}}_{y}
\sum_{i\in\mathcal{I}}
\sum_{t\in\mathcal{T}}
\sum_{h\in\mathcal{H}}
N_{ty}L_{ithy},
\qquad
\forall y
\end{aligned}
\label{eq:cps_compliance}
\end{equation}

\end{subequations}

Eq.~\eqref{eq:resource_adequacy} requires Effective Load Carrying Capability (ELCC)-adjusted accredited capacity to satisfy annual net peak load plus the planning reserve margin. We assume that power imports into Massachusetts are deducted from demand; therefore, both \(\overline{L}_{y}\) and \(L_{ithy}\) are defined on a net-load basis. The nonnegative variable \(p^{\mathrm{RA}}_{y}\) represents any remaining capacity shortfall. Eq.~\eqref{eq:co2_cap} limits annual emissions from RGGI-covered generators to \(\overline{E}^{\mathrm{CO_2}}_{y}\), with the associated allowance cost included in Eq.~\eqref{eq:penalty_cost}. Eq.~\eqref{eq:cps_compliance} enforces annual CPS compliance by requiring multiplier-adjusted CPEC production from eligible battery discharge, together with any shortfall covered by ACP \(e^{\mathrm{CPS}}_{y}\), to meet the applicable minimum standard multiplied by annual Massachusetts net load.

\begin{table}[b]
\vspace{-2em}
\caption{Hourly CPS Multipliers Used in the Model}
\begin{center}
\setlength{\tabcolsep}{3.5pt}
\renewcommand{\arraystretch}{0.95}
\begin{tabular}{|l|l|l|c|}
\hline
\textbf{Period}
& \textbf{Calendar Dates}
& \textbf{Eligible Hours}
& \(\boldsymbol{M^{\mathrm{CPS}}_{thy}}\) \\
\hline
Spring
& Mar.~1--May~14
& 5:00--9:00 P.M.
& 1 \\
\hline
Summer
& May~15--Sep.~14
& 4:00--8:00 P.M.
& 4 \\
\hline
Fall
& Sep.~15--Nov.~30
& 4:00--8:00 P.M.
& 1 \\
\hline
Winter
& Dec.~1--Feb.~28/29
& 4:00--8:00 P.M.
& 4 \\
\hline
Monthly peak
& One hour/month
& Actual system peak hour
& 25 \\
\hline
Other periods
& --
& --
& 0 \\
\hline
\end{tabular}
\label{tab:cps-multipliers}
\end{center}
\end{table}

\section{Case Study}\label{sec:case_study}
\subsection{Data Sources and Assumptions}\label{subsec:data_sources}

The Massachusetts system is represented by three zones: NEMA, SEMA, and WCMA. Existing generation and storage capacities are obtained from the ISO-NE CELT forecast \cite{ISO-NE_CELT}, with resources operating by 2026 treated as existing capacity. Candidate resources comprise zonal four-hour batteries and offshore wind projects identified from the ISO-NE interconnection queue \cite{iso_ne_interconnection_queue}. Candidate capacities are continuously divisible to represent aggregated additions. The directional interzonal transfer limits shown in Fig.~\ref{fig:system_rep} are based on ISO-NE
interface capability assumptions \cite{iso_ne_2026_ttc}.

Hourly zonal demand is constructed from the 2025 Massachusetts load profile \cite{isone_zonal_information} and scaled over 2026--2030 using annual load-growth projections \cite{ISO-NE_LoadForecast}. Assumed imports into Massachusetts are deducted from gross demand; therefore, hourly demand and coincident peak load are defined on a net-load basis. Consistent with the modeled system boundary, annual net load is also used as a proxy for CPS-obligated retail electricity sales. Hourly solar, land-based wind, and offshore-wind availability factors are derived from weather-based profiles \cite{pfenninger2018renewables}. The full 8760-hour chronology is retained in each year to capture seasonal peaks, renewable variability, and intertemporal storage operation.

Technology costs and operating characteristics are based on national technology cost projections \cite{mirletz2024annual}. Capital costs are annualized over the assumed technology lifetimes using a 6\% discount rate, and fixed and variable O\&M costs are expressed in a common dollar year. Four-hour batteries have charging and discharging efficiencies of 95\%, a variable cycling cost of \$3/MWh, and technology-specific ELCC values derived from the regional resource adequacy study \cite{ge_2022_elcc_isone}. The resource adequacy requirement applies an 11\% planning reserve margin to Massachusetts net peak load \cite{nerc_2022_ltra}. The annual RGGI emissions cap is set to 6.17 million tons, with an allowance price of \$27.60/ton \cite{rggi_website}, while the CPS ACP is set to \$65/CPEC \cite{mass_doer_cps}.

The CPS minimum standards are reported in Table~\ref{tab:cps-scenarios}. The NTRM for newly built storage is applied separately and is set to 1 in the pre-May scenario and 2 in the post-May scenario. The hourly CPS multipliers are summarized in Table~\ref{tab:cps-multipliers}. Eligible battery discharge receives the applicable seasonal multiplier during designated business-day peak periods and a multiplier of 25 during the actual monthly system peak hour; all other discharge receives no CPEC credit \cite{mass_doer_cps}.

The model omits unit commitment, start-up and minimum-generation constraints, ancillary-service revenues, endogenous imports, and resource retirements. Existing units remain available throughout the horizon, and each zonal battery represents an aggregated fleet that may charge and discharge simultaneously. These assumptions isolate the capacity-expansion and dispatch effects of the CPS policy packages rather than reproduce all ISO-NE market operations.

\subsection{Numerical Results}
\label{sec:results}

Both the \emph{pre-May} and \emph{post-May} scenarios satisfy the annual CPS requirements without ACP payments. Resource adequacy shortfall and load shedding are also zero in every year. No candidate offshore wind capacity is selected because its higher modeled cost is not justified in the absence of unserved energy or binding resource adequacy constraints. The comparison is therefore driven primarily by the battery capacity and operation required under the two scenarios. Table~\ref{tab:main-results} summarizes the cumulative outcomes.

\subsubsection{Deployment Burden and Certificate Leverage}
\label{subsec:deployment-results}

\begin{figure}[t]
\vspace{-1em}
    \centering
    \includegraphics[width=0.38\textwidth]{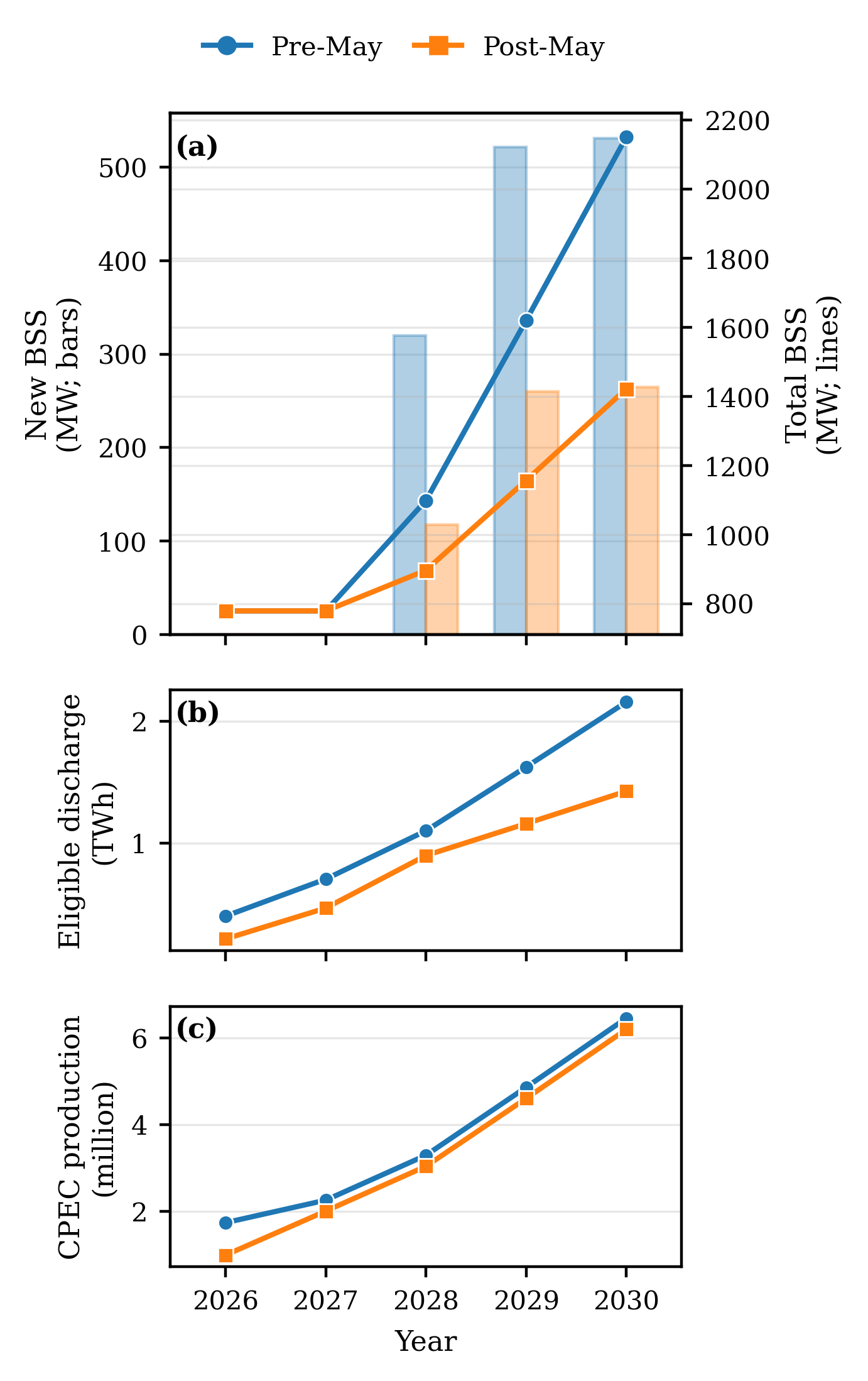}
    \vspace{-1em}  
    \caption{Modeled BSS deployment and CPS response under the \emph{pre-May} and \emph{post-May} scenarios: (a) annual additions and total capacity, (b) CPS-eligible discharge, and (c) CPEC production.}
    \label{fig:storage-response}
    \vspace{-1em}    
\end{figure}

Fig.~\ref{fig:storage-response} compares BSS deployment, physical discharge, and CPEC production. As shown in Fig.~\ref{fig:storage-response}(a), investment begins in 2028 under both scenarios, but the \emph{post-May} scenario requires substantially less capacity. The \emph{pre-May} scenario adds 319.58, 521.54, and 530.87~MW in 2028--2030, yielding 1.372~GW of cumulative additions and 2.151~GW of total BSS capacity in 2030. The \emph{post-May} scenario adds 117.54, 260.39, and 265.05~MW, yielding 642.97~MW of cumulative additions and 1.422~GW of total capacity. Cumulative additions and the maximum annual addition therefore decrease by 53.1\% and 50.1\%, respectively.

\begin{table}[htbp]
\caption{Cumulative Modeled Outcomes, 2026--2030}
\begin{center}
\setlength{\tabcolsep}{3pt}
\renewcommand{\arraystretch}{1.08}
\begin{tabular}{|l|r|r|r|}
\hline
\textbf{Metric}
& \textbf{Pre-May}
& \textbf{Post-May}
& \textbf{Change} \\
\hline
New BSS capacity (MW)
& 1372.0
& 643.0
& $-53.1\%$ \\
\hline
Total BSS capacity in 2030 (MW)
& 2151.0
& 1422.0
& $-33.9\%$ \\
\hline
Eligible BSS discharge (TWh)
& 6.0
& 4.2
& $-30.4\%$ \\
\hline
CPEC production (million)
& 18.6
& 16.8
& $-9.5\%$ \\
\hline
CPS-period fossil generation (TWh)
& 10.0
& 11.1
& $+11.7\%$ \\
\hline
Modeled system cost (\$B)
& 6.6
& 6.4
& $-3.7\%$ \\
\hline
CO$_2$ emissions (Mt)
& 30.8
& 30.8
& $<0.1\%$ \\
\hline
\end{tabular}
\label{tab:main-results}
\end{center}
\end{table}

Figs.~\ref{fig:storage-response}(b) and~\ref{fig:storage-response}(c) show that physical discharge decreases more than certificate production. Cumulative CPS-eligible discharge falls from 5.986 to 4.165~TWh, or 30.4\%, whereas CPEC production falls from 18.613 to 16.848 million, or 9.5\%. The higher NTRM under the \emph{post-May} scenario increases the certificate yield of qualifying new storage, allowing CPS compliance with less installed capacity and physical discharge.

\subsubsection{Dispatch Response Across CPS Multipliers}
\label{subsec:dispatch-results}

\begin{figure}[b]
\vspace{-2em}
    \centering
    \includegraphics[width=0.35\textwidth]{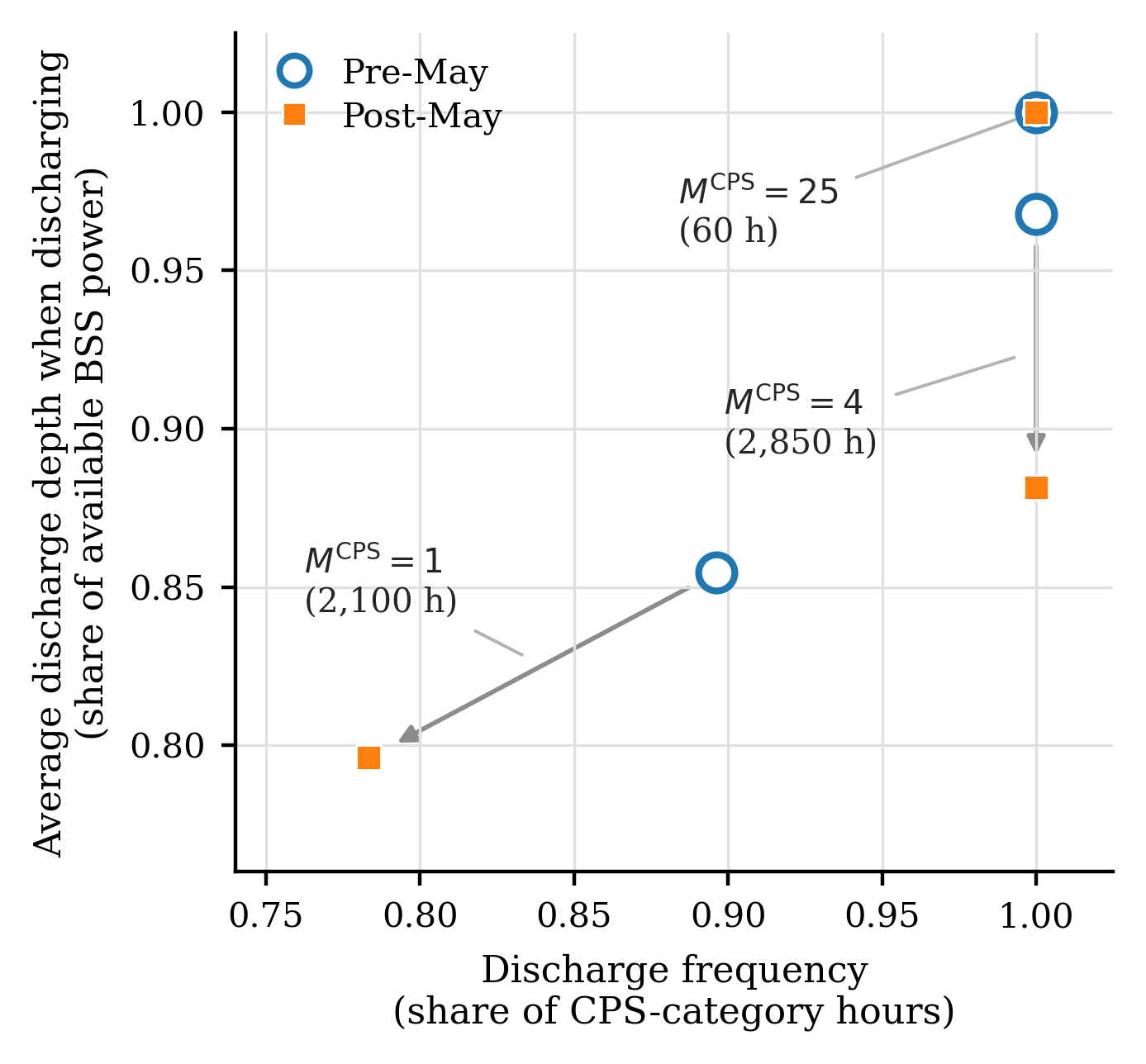}
    \vspace{-1em}
    \caption{BSS discharge frequency and average discharge depth by CPS multiplier under the \emph{pre-May} and \emph{post-May} scenarios. Arrows indicate the operational shift after recalibration.}
    \label{fig:multiplier-response}
\end{figure}

To distinguish installed-capacity effects from operational targeting, Fig.~\ref{fig:multiplier-response} compares BSS discharge frequency and average discharge depth across CPS multiplier categories. Discharge frequency is the share of category hours with positive aggregate BSS discharge, while discharge depth is the average output during those hours as a share of available BSS power capacity.

Both scenarios preserve full discharge frequency for $M^{\mathrm{CPS}}=4$ and 25. At $M^{\mathrm{CPS}}=25$, BSS output also reaches the full available capacity under both scenarios. The main differences occur at lower multipliers. For $M^{\mathrm{CPS}}=1$, frequency decreases from 0.896 in the \emph{pre-May} scenario to 0.783 in the \emph{post-May} scenario, while average depth decreases from 0.854 to 0.796. For $M^{\mathrm{CPS}}=4$, average depth decreases from 0.968 to 0.882. The \emph{post-May} scenario therefore maintains the strongest response during the monthly peak hour but produces less frequent and shallower discharge in other eligible periods.

\subsubsection{Cost and Physical Clean-Peak Tradeoffs}
\label{subsec:tradeoff-results}

\begin{figure*}[t]

    \centering
    \includegraphics[width=0.8\textwidth]{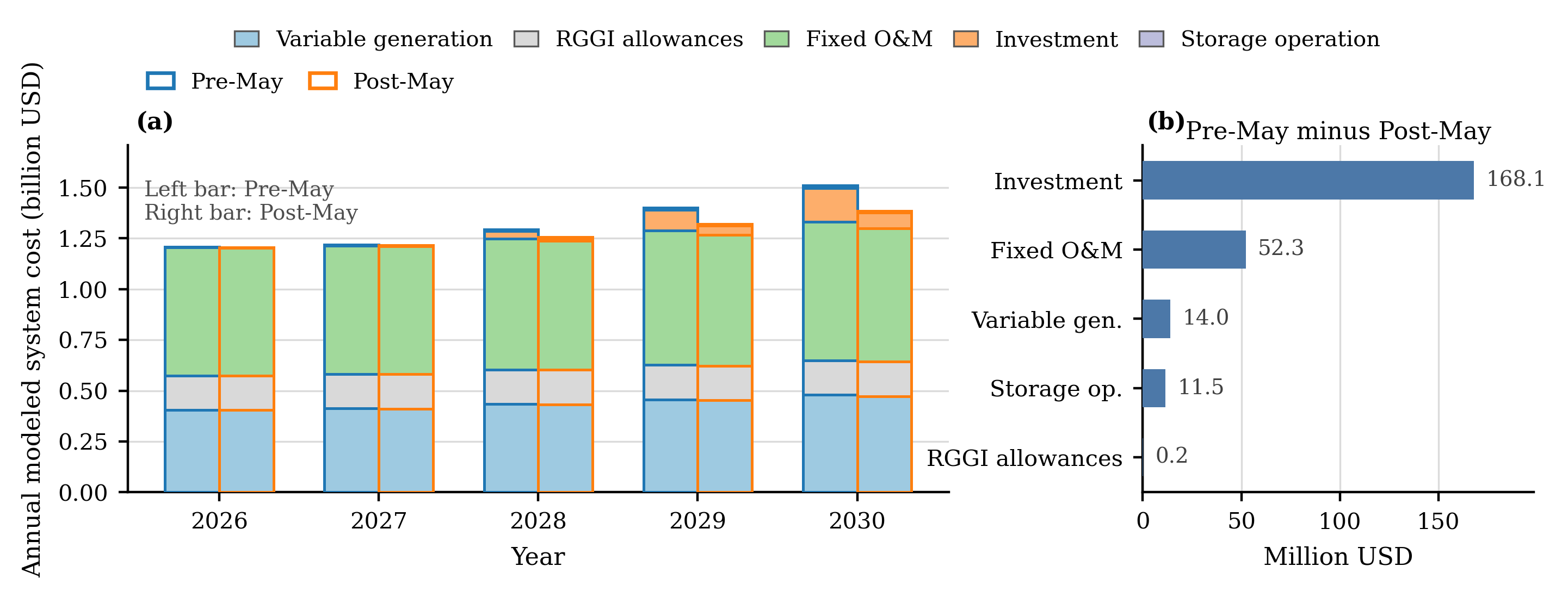}
    \vspace{-1em}  
    \caption{Modeled system-cost comparison under the \emph{pre-May} and
\emph{post-May} scenarios: (a) annual cost composition and (b) five-year cost
reduction, measured as \emph{pre-May} minus \emph{post-May}.}
    \label{fig:cost-results}
      \vspace{-1em}  
\end{figure*}

The \emph{post-May} scenario reduces five-year modeled system cost from \$6.631 billion to \$6.385 billion, a savings of \$246.1 million, or 3.7\%. Fig.~\ref{fig:cost-results}(a) shows that the cost difference widens after BSS investment begins in 2028. As shown in Fig.~\ref{fig:cost-results}(b), lower investment accounts for \$168.1 million, or approximately 68\% of the total savings. Lower fixed O\&M, variable-generation, and storage operation costs contribute \$52.3 million, \$14.0 million, and \$11.5 million, respectively, whereas RGGI allowance costs change by only \$0.2 million.

The lower-cost outcome is accompanied by weaker physical clean-peak performance. Cumulative fossil generation during CPS hours increases from 9.955 to 11.120~TWh, or 11.7\%, while five-year CO$_2$ emissions change by only $-0.02\%$. Because the RGGI cap binds from 2027 onward under both scenarios, the additional CPS-period fossil generation is offset by redispatch in other hours. Thus, RGGI constrains annual emissions, whereas the CPS primarily affects the timing of storage and fossil operation.

Overall, the \emph{post-May} scenario provides a lower-cost compliance pathway in the modeled system. It preserves full CPS compliance and strong BSS response during the highest-weighted hours while reducing cumulative and maximum annual storage additions. The tradeoff is that certificate production is maintained more effectively than physical discharge and fossil displacement during CPS periods. These results reflect the combined changes in the minimum standards and candidate NTRM; they quantify modeled system effects rather than project delivery under permitting, procurement, equipment, or interconnection constraints.

\section{Conclusion}
\label{sec:conclusion}

This paper quantifies how Massachusetts' 2026 CPS recalibration changes storage investment and system operation. The \emph{post-May} scenario maintains full CPS compliance while reducing new BSS additions from 1.372 to 0.643~GW and lowering modeled system cost by 3.7\%. However, CPS-eligible discharge falls by 30.4\%, and fossil generation during CPS hours rises by 11.7\%, even though modeled emissions remain nearly unchanged under the binding RGGI cap.

The results therefore identify a clear tradeoff. The \emph{post-May} scenario reduces the modeled storage capacity requirement and total system cost, but preserves certificate production more effectively than physical clean-peak performance. Because the minimum standards and NTRM change together, the reported outcomes reflect their combined effect. Overall, the recalibrated CPS provides a less demanding compliance pathway, but achieves less storage deployment and fossil displacement during CPS hours.

\renewcommand{\bibfont}{\small}
\bibliographystyle{IEEEtran}
\bibliography{Ref}

\end{document}